\documentclass[twocolumn,amsmath,showkeys,prb,citation]{revtex4-1}

\usepackage{dcolumn}% Align table columns on decimal point
\usepackage{bm}% bold math
\usepackage[T1]{fontenc}
\usepackage{amsmath}
\usepackage{graphicx}
\usepackage{graphics}
\usepackage{epstopdf}

\usepackage{hyperref}
\begin{document}
\title{First-principles study of PbTiO$_3$ under uniaxial strains and stresses}

\author{Henu Sharma$^{1,2,3}$}
\author{Jens Kreisel$^{3,4}$}
\author{Philippe Ghosez$^{1}$}

\affiliation {$^1$ Physique Th\'eorique des Mat\'eriaux, Universit\'e de Li\`ege, B-4000 Sart Tilman, Belgium}
\affiliation{$^2$ Lab. Mat\'eriaux et G\'enie Physique, Grenoble INP-CNRS, Grenoble, France}
\affiliation{$^3$ Science and Analysis of Materials, Gabriel Lippmann Public Research Center, 41 Rue du Brill, L-4422 Belvaux, Luxembourg}
\affiliation{$^4$ Physics and Materials Science Research Unit, University of Luxembourg, 41 Rue du Brill, L-4422 Belvaux, Luxembourg}

\begin{abstract}

The behavior of PbTiO$_3$ under uniaxial strains and stresses is investigated from first-principles calculations within density functional theory. We show that irrespectively of the uniaxial mechanical constraint applied, the system keeps a purely ferroelectric ground-state, with the polarization aligned either along the constraint direction ($FE_z$ phase) or along one of the pseudo-cubic axis perpendicular to it ($FE_x$ phase). This contrasts with the cases of isotropic or biaxial mechanical constraints for which novel phases combining ferroelectic and antiferrodistortive motions have been previously reported. Under uniaxial strain, PbTiO$_3$ switched from a $FE_x$ ground state under compressive strain to $FE_z$ ground-state under tensile strain, beyond a critical strain $\eta_{zz}^c \approx +1$\%. Under uniaxial stress, PbTiO$_3$ exhibits either a $FE_x$ ground state under compression ($\sigma_{zz} < 0$) or a $FE_z$ ground state under tension ($\sigma_{zz} > 0$). Here, however, an abrupt jump of the structural parameters is also predicted under both compressive and tensile stresses at critical values $\sigma_{zz} \approx$ $+2$ GPa and $- 8$ GPa. This behavior appears similar to that predicted under negative isotropic pressure and might reveal practically useful to enhance the piezoelectric response in nanodevices.
\end{abstract}

\maketitle

\section{Introduction}

$AB$O$_3$ perovskites form a very important class of functional materials that can exhibit a broad range of properties (e.g. superconductivity, magnetism, ferroelectricity, multiferroism, metal-insulator transitions ...) within small distortions of the prototype cubic structure. Amongst them, PbTiO$_3$ is a prototypical ferroelectric compound and also one of the parent components of Pb(Zr,Ti)O$_3$ solid solution (PZT), which is the most widely used piezoelectrics \cite{lines77a}.

Bulk PbTiO$_3$ crystallizes at high temperature in the paraelectric $Pm\bar{3}m$ cubic structure. Under cooling, it then undergoes at 760 K a structural phase transition to a ferroelectric phase of $P4mm$ symmetry. At room temperature, it possesses a large spontaneous polarization $P_s \approx 80$ $\mu$C/cm$^2$. Contrary to BaTiO$_3$ and KNbO$_3$ that exhibit additional ferroelectric transitions to phases of orthorhombic and rhombohedral symmetries, PbTiO$_3$ remains tetragonal down to zero Kelvin, a feature that was assigned to its large $c/a$ ratio \cite{Cohen92}. As revealed from the inspection of the phonon dispersion curves of its cubic phase \cite{Ghosez99}, on top of its ferroelectric instability, PbTiO$_3$ also develops strong antiferrodistortive (AFD) instabilities, associated to rotations (tilts) of the oxygen octahedra. Although, these AFD instabilities are suppressed by the appearance of the dominant FE motions, they nevertheless constitute important hidden instabilities that can significantly affect its physical and structural properties. For instance, it was recently highlighted theoretically that AFD motions shift down the ferroelectric phase transition temperature of PbTiO$_3$ by few hundreds of Kelvin \cite{Wojdel-13}. Also, although they do not naturally appear in bulk, AFD motions can condense at the PbTiO$_3$ surface \cite{Bungaro05} where the FE-AFD competition is modified. 

In ABO$_3$ compounds, FE and AFD instabilities are highly sensitive to mechanical constraints as strains and stresses, that can thus be used in practice to tune the phase transition temperatures and the multifunctional properties \cite{lines77a}. Under increasing isotropic pressure, the ferroelectric instability is so well-known to disappear quickly, an intrinsic feature that has to be properly handled when doing first-principles calculations within the local density approximation that tends to underestimate systematically bond-lengths and unit-cell volumes \cite{Rabe-Ghosez}. Unexpectedly, in PbTiO$_3$, Kornev {\it et al.} \cite{kornev05a} have shown that, although ferroelectricity is indeed progressively suppressed at low isotropic pressure, it reappears at ultrahigh pressure, a feature also predicted in BaTiO$_3$ \cite{ericphilippe}. Following this, the phase diagram of PbTiO$_3$ under isotropic pressure has been recently reinvestigated by Janolin {\it et al.} \cite{kreisel08a}: they highlighted a complex sequence of phases accommodating pressure through mechanisms involving not only the reentrance of ferroelectricity but also oxygen octahedra tilting, which are known to be favored at smaller volumes.

Engineering ferroelectricity through biaxial epitaxial strain in ABO$_3$ thin films has also attracted much attention over the last decade \cite{Review-Rabe}. Thanks to the advances in the deposition of coherent epitaxial films of complex oxides, \cite{triscone} it has become possible to impose strains of the order of 4\% or even larger to thin-film perovskites. It is now well understood that the substrate-induced biaxial strain has a strong bearing on the ultimate behavior of ferroelectric thin films \cite{Review-Javier}. Prototypical demonstrations of this include the strong amplification of ferroelectricity in strained BaTiO$_3$ \cite{Eom} or the possibility to achieve room-temperature ferroelectricity in strained SrTiO$_3$ \cite{gopalan,haeni}. Such behaviors were predicted by Landau theory \cite{pertsevm} and further analyzed from first-principles investigations \cite{dieguez}. Strain engineering of ferroelectricity was also considered as a promising route to convert paraelectric magnets into multiferroics \cite{Review-Varignon} like for instance in CaMnO$_3$ \cite{Battacharjee,Gunter} and is not restricted to perovskites \cite{Bousquet-AO}. Beyond acting simply on the ferroelectric mode as initially targeted, strain engineering revealed also useful to tune the competition with other instabilities and get novel unexpected phases like in BiFeO$_3$ under either in-plane compressive \cite{BFO-compressive} or tensile \cite{BFO-tensile} strains, or in EuTiO$_3$ \cite{Fennie-ETO}, in combination with magnetism. In PbTiO$_3$ it was predicted from first-principles that, while compressive strain will favor the $P4mm$ ferroelectric phase and amplify the spontaneous polarization, tensile epitaxial strain should favor an $Ima2_1$ phase \cite{ericthesis} combining in-plane FE polarization (along the [110] direction) and in-plane AFD oxygen rotations ($a^-a^-c^0$ in Glazer's notations \cite{Glazer,Howard}).

While the effect of isotropic and biaxial mechanical constraints on the ferroelectric properties has been widely investigated, our study is motivated by the little existing understanding of the effect of uniaxial strain and stress. Regarding PbTiO$_3$, we can cite a recent study by Yifeng Duan {et al.} \cite{yifeng}, but we note that the authors did not consider the possible interplay of ferroelectricity with AFD motions and, moreover, as it will appear clearer below, their conclusions are biased by the fact that they restricted to a particular phase. Within our study, it is interesting to explore if uniaxial pressure leads to a suppression of FE in favour of AFD distortions or if FE structures are favoured at any uniaxial strain or stress state. Also, we wish to explore, if uniaxial strain /stress leads to new structures in PbTiO$_3$, based on the fact that perovskites with competing FE and AFD instabilities can show a multitude of structures under deformation, as observed under biaxial strain for BiFeO$_3$ \cite{BFO-compressive} or under hydrostatic pressure not only for PbTiO$_3$ but also for BiFeO$_3$ \cite{Guennou-11} or BiMn O$_3$ \cite{Guennou-14}. Here we perform first-principles calculations within density functional theory in order to determine the ground state of PbTiO$_3$ under uniaxial strains and stresses, searching for potential transitions to unexpected phases.

\section{Technical Details}

Our first-principles calculations have been performed in the framework of density functional theory (DFT) as implemented in the ABINIT package \cite{gonze02a,matthieu,gonze09a}. We did calculations using both (i) the local density approximation (LDA)\cite{Teter-LDA} and extended norm-conserving Teter pseudopotentials~\cite{teter93a} and (ii) the generalized gradient approximation with the functional proposed by Wu and Cohen (GGA-WC) \cite{wu06a} and optimized RRK pseudopotentials \cite{RRK} generated with OPIUM code \cite{OPIUM}. In both cases, semi-core states were treated as valence electrons, considering explicitly the following levels in the calculation: {5\it{s}}, {5\it{p}} and {6\it{s}} for the Pb atom, {3\it{s}}, {3\it{p}}, {3\it{d}} and {4\it{s}} for the Ti atom and {2\it{s}} and {2\it{p}} for the O atom. The wavefunction was expanded on a plane-wave basis set. Convergency was reached using a plane-wave energy cutoff of 45 hartrees. 

In the 5 atoms perovskite $AB$O$_3$ unit cell, a Monkorsh-Pack mesh of 6$\times$6$\times$6 {\it k} points was used to sample the Brillouin zone. When condensing the AFD instabilities, we considered either a 20-atom supercell corresponding to $\sqrt{2}$a$_0$= a=b, and c= 2a$_0$, and a sampling of 6$\times$6$\times$4 {\it k} points or, for the {\it Cmcm} phase, a 40-atom supercell corresponding to 2a$_0$, 2a$_0$, and 2a$_0$ and a sampling of 4$\times$4$\times$4 {\it k} points. We explicitly checked that the relative energy of the different phases is well converged and independent of the choice of the supercell. Structural relaxations were performed until the forces were smaller than $10^{-7}$ hartrees/bohr and stresses are smaller than 10$^{-8}$ hartrees/bohr$^3$. The vibrational properties, Born effective charges and dielectric tensors were calculated using the density functional perturbation theory (DFPT) \cite{gonze97b}. The spontaneous polarization were computed making use of the Berry phase formalism \cite{KingSmith}.

\begin{figure}[htbp!]
\centering
\includegraphics[scale=.35]{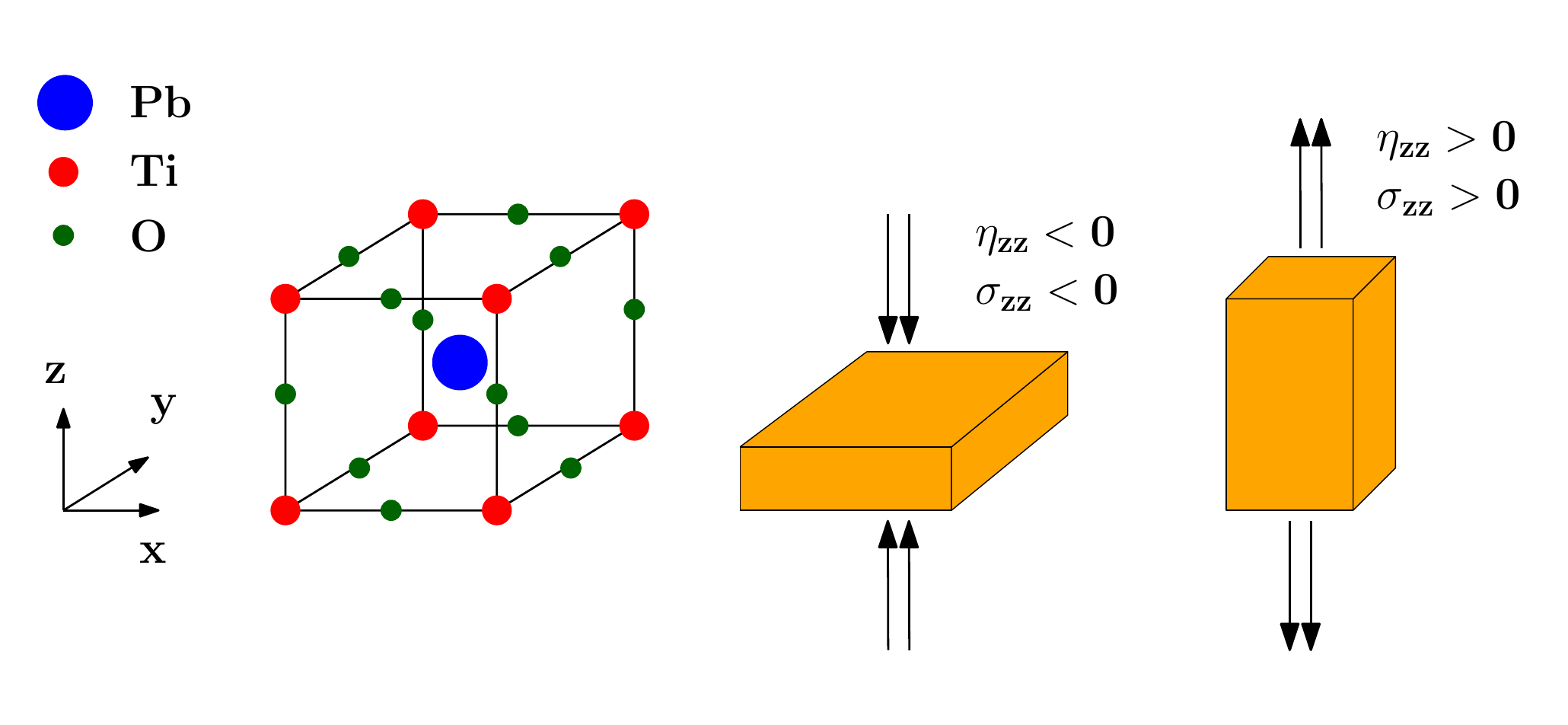}
\caption{{(Color Online) Cubic perovskite structure of PbTi$O_3$, with Ti atom at the origin. Pb atoms are located at the center (in blue), Ti atoms at the corners (in red), and O atoms at the middle of the edges (in green) . The uniaxial mechanical constraint (fixed strain $\eta_{zz}$ or fixed stress $\sigma_{zz}$) is applied along the $z$-axis .}}
\label{fig:ABO}
\end{figure}

In what follows, we consider that the $x$, $y$ and $z$ cartesian axis are aligned respectively with the cell vectors $a$, $b$ and $c$ of the reference cubic structure. Then, the uniaxial constraint is always applied along the $z$-direction as illustrated in Fig. \ref{fig:ABO}. Both fixed uniaxial strain and fixed uniaxial stress conditions will be considered. 

To label the ferroelectric and antiferrodistortive motions compatible with a given space group, we use ``extended'' Glazer's notations in which the superscripts refer as usual to the rotation pattern and a subscript $P$ is added to identify the direction(s) along which a polarization can develop. When reporting phonon labels, we consider that the Ti atom is at the origin.

\section{Bulk structure}

First, we reinvestigate the highly-symmetric cubic perovskite structure of PbTiO$_3$. In this cubic phase, the atomic positions are fixed by symmetry and the only structural parameter to be relaxed is the lattice constant $a_0$. Our relaxed lattice constants a$_0^{LDA} =$ 3.880 \AA$\,$ and a$_0^{GGA} =$ 3.933 \AA$\,$ are comparable to previous calculations (a$_0^{LDA}$ = 3.874 \AA \cite{yifeng}) and in satisfactory agreement with experimental data (a$_0^{EXP}$ = 3.97 \AA \cite{shirane56a}). As expected, the LDA tends to underestimate the experimental lattice constant that is better reproduced at the GGA-WC level.

The calculated phonon dispersion curves of cubic PbTiO$_3$ (not shown here) are also in agreement with previous literature \cite{Ghosez99}. They show two main phonon instabilities: (i) a zone-center FE unstable mode $\Gamma^{4-}$ ($F_{1u}$) at 109i (151i) cm$^{-1}$ in LDA (GGA-WC) corresponding at a polar displacement of cations against the oxygen and (ii) a zone-boundary AFD unstable mode $R^{4+}$ at 98i (79i) cm$^{-1}$ in LDA (GGA-WC) corresponding to rotations of the oxygen octahedra, with consecutive octahedra along the rotation axis moving anti-phase ($a^-$ in Glazer's notations). As usual in perovskites, we notice that the AFD instability at R-point propagates to the M-point through a $M^{3+}$ mode at 73i (53i) cm$^{-1}$ in LDA (GGA-WC) where consecutive octahedra move in-phase ($a^+$ in Glazer's notations). The main difference between LDA and GGA-WC results comes from the smaller LDA volume that favors the AFD instabilities and reduces the FE instability.

\begin{table}[h]
\begin{center} \footnotesize
\begin{tabular}{l  c c c c c l|c|c|c|c|c|} \hline \hline 
Phase   &   &   Unit cell    & & Energy & Distortion(s) \\ 
     &               a & b  & c &      $\Delta$E     & Angle or $P_s$  \\
     &               (\AA) & (\AA)  & (\AA) &      (meV/f.u.)     & ($^{\circ}$ or $\mu$C/cm$^2$) \\
		\hline
$Pm\bar{3}m$    & $3.880$ & $3.880$ & $3.880$ & 0 & --\\ 
(a$^0$a$^0$a$^0$) & $(3.935)$ & $(3.935)$ & $(3.935)$ & (0) & -- \\ 
{\it Exp.}\cite{shirane56a}   & 3.97   & 3.97   &3.97 &-- & -- \\
\hline 	
$P4mm$  &  $3.863$ & $3.863$ & $3.975$ & -36.70   & $P_s=70$ \\ 
(a$^0$a$^0$c$^0_P$)&  $(3.880)$ & $(3.880)$ & $(4.243)$ & (-83.27) & ($P_s=97$) \\ 
{\it Exp.}\cite{} &  {\it $3.880$} & {\it$3.880$} & {\it$4.155$} & -- & {\it$P_s= 81$}  \\ 
%{\it LGD}\cite{} &  {\it$xxx$} & {\it$xxx$} & {\it$xxx$} & - & {\it$P_s= 76$}  \\
\hline
$Amm2$  &  $3.912$ & $3.912$ & $3.865$ & -31.60   & $P_s=63$ \\ 
(a$^0_P$a$^0_P$c$^0$)&  $(3.999)$ & $(3.999)$ & $(3.901)$ & (-62.90) & ($P_s=75$) \\ 
\hline
$R3m$  &  $3.895$ & $3.895$ & $3.895$ & -30.02   & $P_s=62$ \\ 
(a$^0_P$a$^0_P$a$^0_P$)&  $(3.962)$ & $(3.962)$ & $(3.962)$ & (-58.25) & ($P_s=71$) \\ 
\hline
$P4/mbm$     &  $5.477$ & $5.477$ & $7.783$    & -3.20 &    $\phi^+=$ 4.13 \\
(a$^0$a$^0$c$^+$) &   $(5.558)$  & $(5.558)$ & $(7.880)$ & (-1.06) & ($\phi^+=$ 3.09)\\ 
\hline
$I4/mcm$   &  $5.470$ & $5.470$ & $7.797$ & -10.80 & $\phi^-=$ 5.62 \\ 
(a$^0$a$^0$c$^-$) &  $(5.552)$ & $(5.552)$ & $(7.891)$ & (-5.00) & ($\phi^-=$ 4.60)\\ 
\hline    
$Imma$   &  $5.480$ & $5.505$ & $7.732$ &  -12.01 &  $\phi^-=$ 4.15 \\ 
(a$^-$a$^-$c$^0$)& $(5.559)$ & $(5.576)$ & $(7.850)$ & (-5.59) & ($\phi^-=$ 3.43)\\ 
\hline  
$R\bar{3}c$   &  $7.756$ & $7.756$ & $7.756$ &     -12.00 &   $\phi^-=$ 3.36 \\ 
(a$^-$a$^-$a$^-$) & $(7.866)$ & $(7.866)$ & $(7.866)$ & (-5.29)  & ($\phi^-=$ 2.7)\\ 
\hline            
\hline
\end{tabular}
\end{center} 
\caption{\small Cell parameters, internal energies and distortion amplitudes of different meta-stable phases of PbTiO$_3$ fully relaxed within the LDA and the GGA-WC (values in brackets). For each phase, we specify the space group and, in brackets, the compatible FE and AFD structural distortion using generalized Glazer's notations (see method Section). The amplitude of the spontaneous polarization ($P_s$) and of the oxygen octahedra rotation angle ($\phi$) are reported when appropriate. For the $Pm\bar{3}m$ and $P4mm$ phases, the experimental parameters (Exp.) are reported for comparison.} 
\label{tab:rotations} 
\end{table} 

From this discussion, it appears that, at the harmonic level, the FE instability is stronger than the AFD ones. However, this does not necessarily imply a FE ground state. In Table \ref{tab:rotations}, we report the energy and structural parameters of different metastable phases resulting from the condensation of the FE mode at $\Gamma$ and AFD modes at R and M points. Both LDA and GGA-WC correctly reproduce the $P4mm$ ferroelectric ground state. We see that in absence of FE instability, PbTiO$_3$ would prefer to develop $a^-$ rotation patterns and would adopt one of the $Imma$ ($a^-a^-c^0$) or $R\bar{3}c$ ($a^-a^-a^-$) phase that both appear nearly degenerated in energy in our calculations (i.e. with a difference of energy smaller than 1 meV/f.u.). The $I4/mcm$ ($a^0a^0c^-$) phase is also very close in energy. In comparison, the $a^+$ rotation pattern is never producing a substantial gain of energy; in line with this, we notice that atomic relaxations in the $Cmcm$ ($a^0b^+c^-$) and $Pbnm$ ($a^-a^-c^+$) symmetries relax back to the $I4/mcm$ and $Imma$ phases respectively, attesting that the appearance of the $a^-$ rotation suppresses the instability associated to $a^+$ motions.
 
\begin{figure}[htbp!]
\centering
\includegraphics[scale=.35]{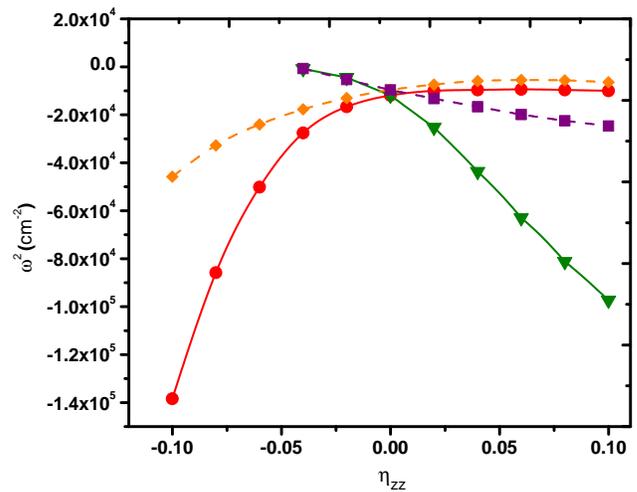}
\caption{{(Color Online) Evolution of the square of the frequency of the FE modes $\Gamma^{3-}$ (green triangles) and $\Gamma^{5-}$ (red circles) and of the AFD modes $A_3^+$ (purple squares) and $A_5^+$ (orange diamonds) with uniaxial strains in the paraelectric $P4/mmm$ phase of PbTiO$_3$, as obtained within the LDA. Similar results have been obtained within the GGA-WC.}}
\label{fig:lowest_optical_phonons}
\end{figure}

Applying uniaxial strain along the $z$-direction and relaxing the lattice constant along the two other directions, while keeping the atoms at their high-symmetry position makes the paraelectric reference unit cell tetragonal, bringing the system from $Pm\bar{3}m$ to $P4/mmm$ symmetry. This splits the triply degenerated $\Gamma^{4-}$ ($F_{1u}$) FE mode into a single $\Gamma^{3-}$ ($A_{1}$) mode and a doubly degenerated $\Gamma^{5-}$ ($E$) mode, polarized respectively along $c$-axis and perpendicularly to it. Similarly the triply degenerated $R^{4+}$ AFD mode is split into a single $A_3^+$ mode and a doubly degenerated $A_5^+$ mode, corresponding respectively to oxygen rotations around the $z$-axis or around the $x$- and $y$-axis. The evolution of the frequencies of these modes with uniaxial strain are reported in Fig. \ref{fig:lowest_optical_phonons}. It appears that while the FE instability is only marginally more unstable than the AFD one at the bulk level, both tensile and compressive uniaxial strains destabilize more strongly one of the FE modes ($\Gamma^{3-}$ under tension and $\Gamma^{5-}$ under compression) than any of the AFD ones. Although limited to the harmonic level, this observation already suggests that the behavior of PbTiO$_3$ under uniaxial mechanical constraints is strongly dominated by the FE instability.

\section{Uniaxial strain}

\begin{figure}[htbp!]
\centering
\includegraphics[scale=.35]{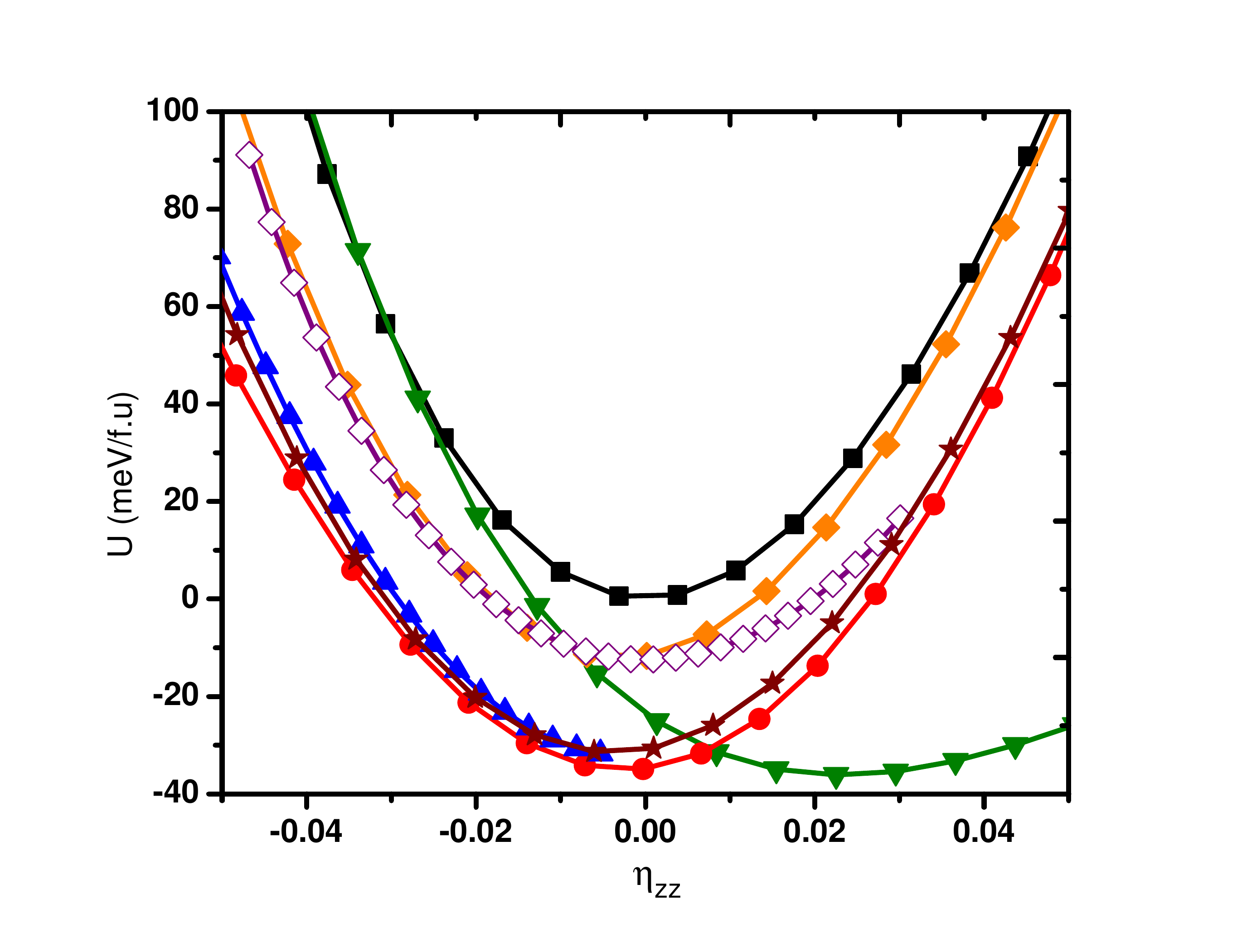}
\includegraphics[scale=.35]{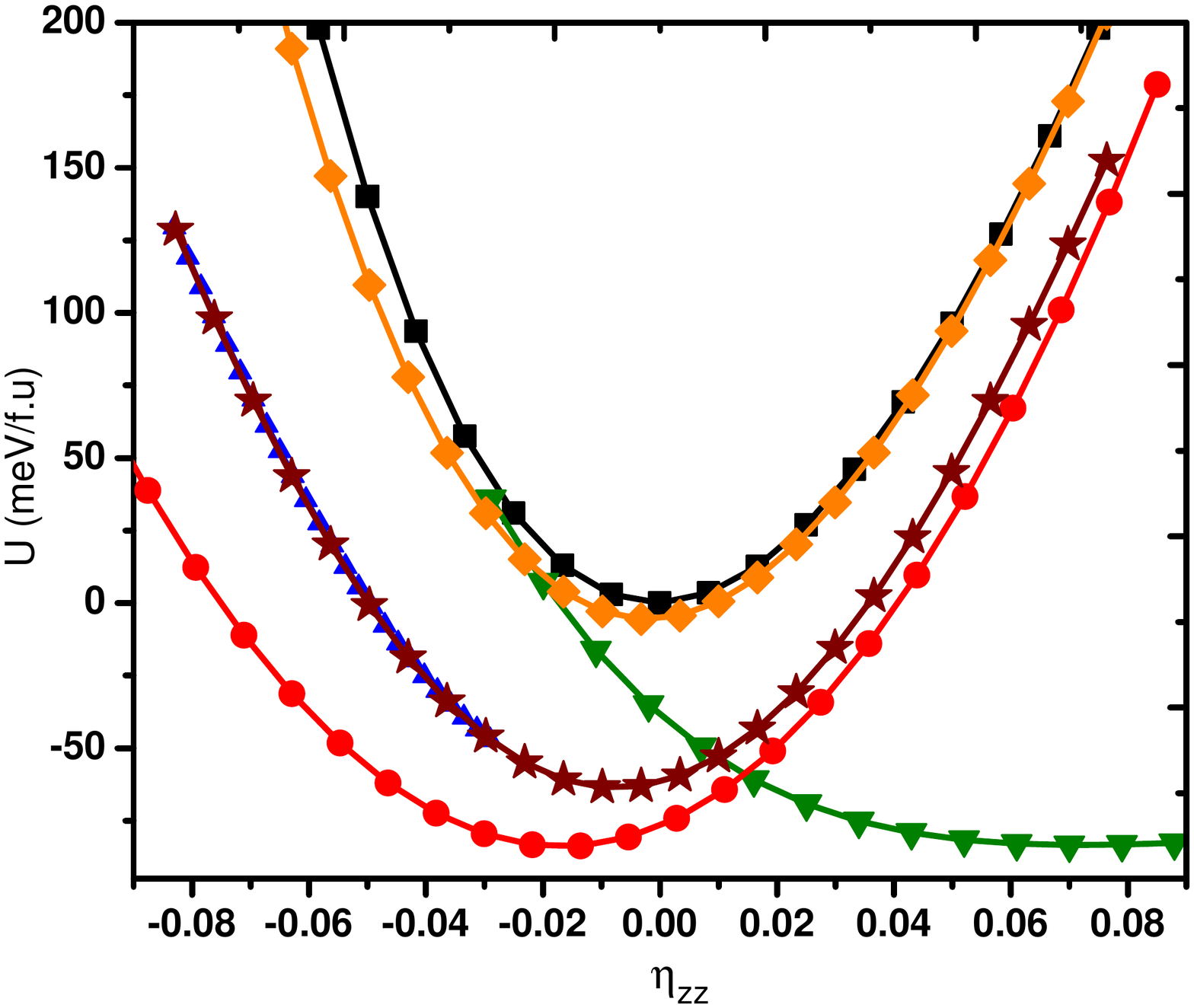}
\caption{{ (Color Online) Internal energy U (meV/f.u.) of different metastable phases of PbTi$O_3$ under uniaxial strain as computed within the LDA (panel a) and the GGA-WC (panel b). The considered phases are the following: PE ($P4/mmm$, black squares), $P_z$ ($P4mm$, green triangles), $P_{xy}$ ($Amm2$, blue triangles), $P_x$ ($Pmm2$, red circles), $AFD_{xy}$ ( orange diamonds), $AFD_{xyz}$ (open purple diamonds) and $AFD_{xy}$+$P_{xy}$ (brown stars). }}
\label{fig:strain_phase}
\end{figure}
		
Let us now focus on the behavior of PbTiO$_3$ under uniaxial strain. The mechanical constraint is applied along the $z$-axis by fixing the $c$ lattice parameter. Then, structural relaxations are performed under different symmetry constraints, in order to compare the stability of different metastable phases for different amplitudes of the strain $\eta_{zz}$. The most stable phase at a given $\eta_{zz}$ is that which minimizes the internal energy $U$. The results obtained in LDA are summarized in Fig. \ref{fig:strain_phase}a and GGA-WC results in Fig. \ref{fig:strain_phase}b. In the following, we will mostly refer to LDA results, unless a further consideration of GGA-WC is pertinent. 

The relaxed $Pm\bar{3}m$ cubic phase of PbTiO$_3$ has a lattice constant $a^{LDA}= 3.880$ \AA$\,$ ($a^{GGA}= 3.935$ \AA) and is chosen as the common reference for both the internal energy ($U=0$) and the strain ($\eta_{zz} = (c -c_0)/c_0$ with $c_0 = a^{LDA}$ or $a^{GGA}$). 

Applying a strain $\eta_{zz}$ to the paraelectric $Pm\bar{3}m$ phase, while keeping the atoms at their high-symmetry positions, brings the system into the $P4/mmm$ symmetry (black squares in Fig. \ref{fig:strain_phase}) . As highlighted in the previous section, this paraelectric (PE) phase is not the ground state: it exhibits different FE and AFD instabilities, the condensation of which will necessarily lower the internal energy. 

\begin{figure}[htbp!]
\centering
\includegraphics[scale=.35]{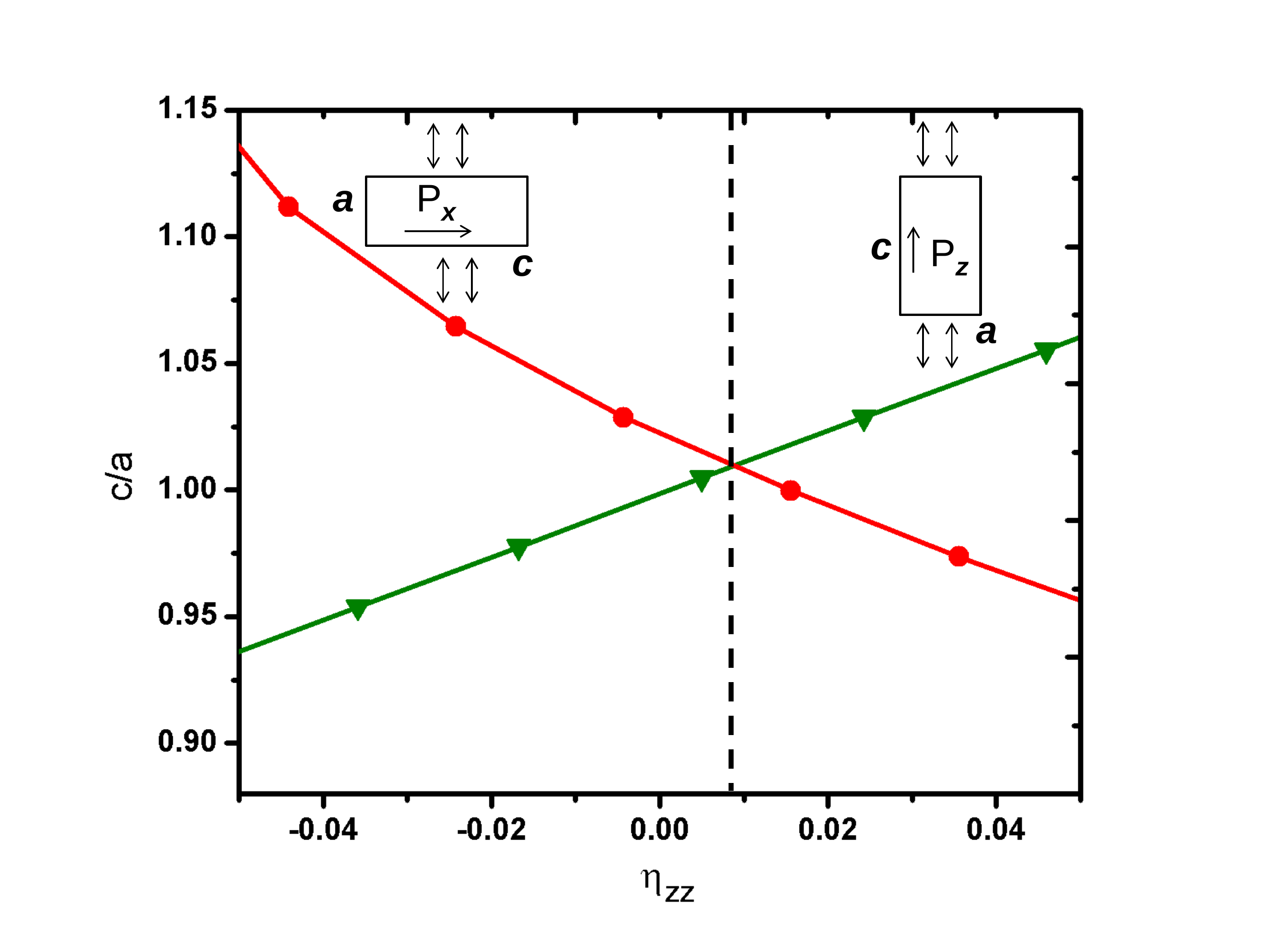}
\caption{{(Color Online) Evolution of the $c/a$ ratio of the relaxed $FE_x$ ($Pmm2$) and $FE_z$ ($P4mm$) phases of PbTiO$_3$ with the uniaxial strain as obtained in LDA. }}
\label{fig:c_over_a}
\end{figure}

Distinct polar phases, with their polar axis aligned along different directions, have then been relaxed. They are labelled $FE_z$ ($P4mm$), $FE_x$ ($Pmm2$) or $FE_{xy}$ ($Amm2$) depending if the polar axis is along the [001], [100] or [110] direction respectively \cite{footnote_1}. We see in Fig. \ref{fig:strain_phase} that $FE_z$ and $FE_x$ curves have their minimum at the same internal energy, respectively for a value of strain associated in LDA (GGA-WC) to $c = 3.975$ \AA$\,$ (4.243 \AA) and $a = 3.863$ \AA$\,$ (3.880 \AA) which correspond to the $c$ and $a$ relaxed lattice constants of the bulk $P4mm$ ground state. The relative position of these two curves is such that the $FE_x$ phase appears to be the most stable for $\eta_{zz} < +0.8$\% while the $FE_z$ phase is favored under tensile strains $\eta_{zz} > 0.8$\%. In Fig. \ref{fig:c_over_a} we have plotted the $c/a$ ratio of the $FE_z$ and $FE_x$ phases: we see that the crossing of the two curves coincide with the change of stability of the two phases, emphasizing that PbTiO$_3$ prefers at each strain the phase that maximizes its $c/a$ ratio. 

\begin{figure}[htbp!]
\centering
\includegraphics[scale=.35]{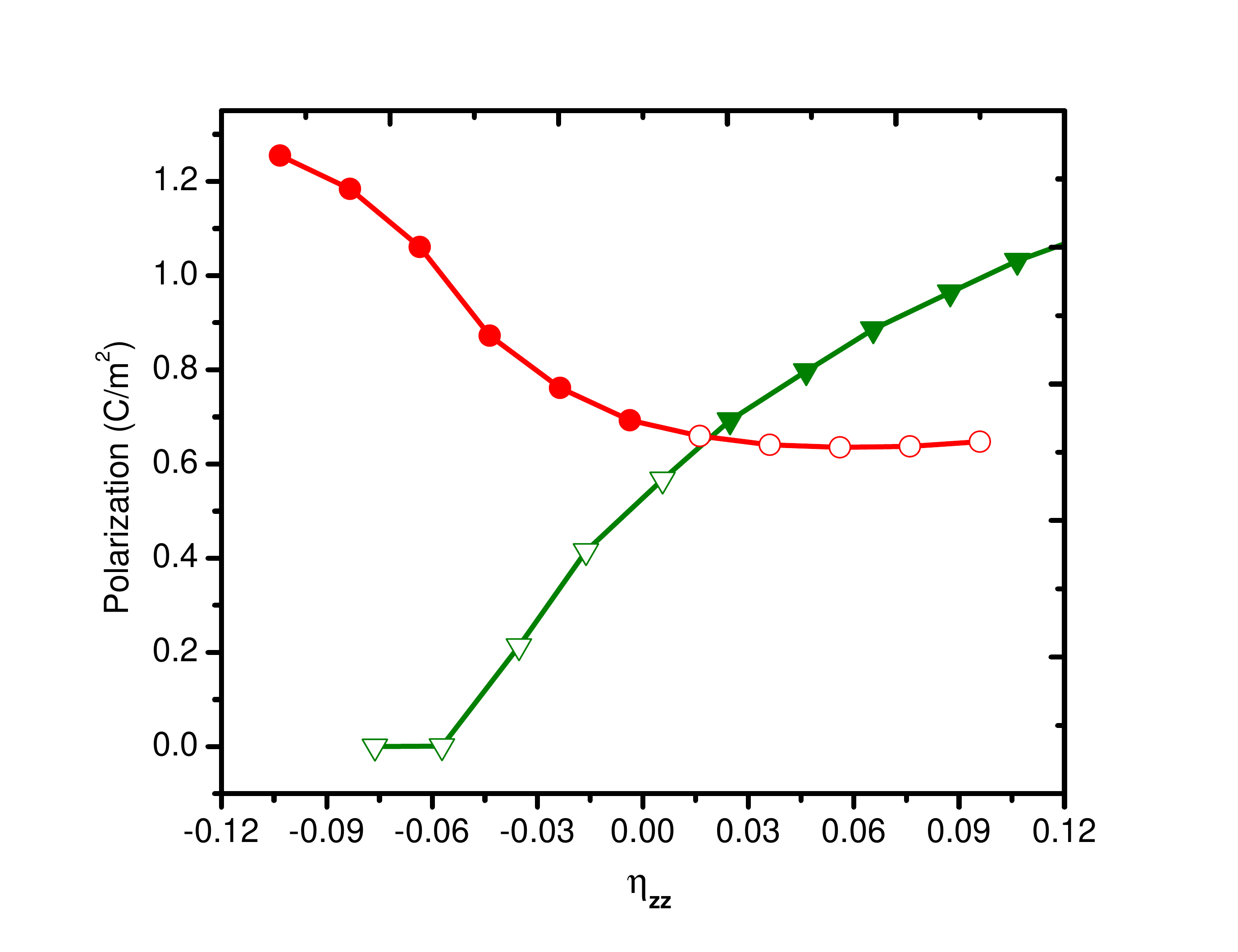}
\includegraphics[scale=.35]{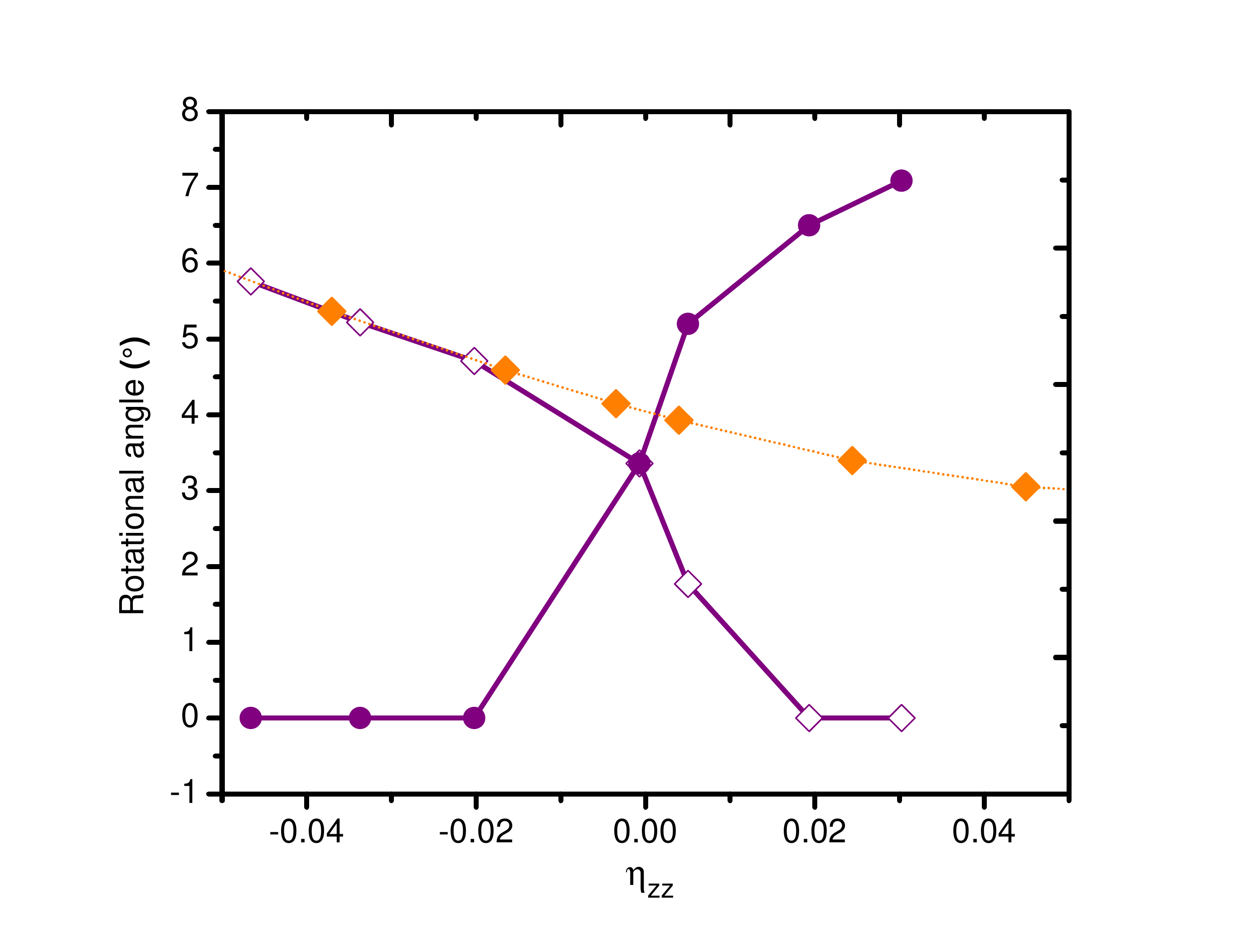}
\caption{{(Color Online) Top: Evolution of the polarization of the $FE_x$ (red circles) and $FE_z$ (green triangles) phases of PbTiO$_3$ with the uniaxial strain, as predicted within the LDA. Filled symbols correspond to the region where the phase is the ground-state \cite{Note-P}. Bottom: Evolution of the rotational angles of the $AFD_{xy}$ and $AFD_{xyz}$ phases of PbTiO$_3$ with the uniaxial strain, as predicted within the LDA. Purple solid circles and open diamonds indicate respectively c$^-$ and a$^-$ rotation angles of the $AFD_{xyz}$ phases and orange filled diamonds indicate a$^-$ rotation angles of the $AFD_{xy}$ phases. }}
\label{fig:Rot}
\end{figure}

We notice also in Fig. \ref{fig:strain_phase} that, contrary to what was proposed in Ref. \cite{yifeng}, the paraelectric configuration is never the most stable. Consistently with that work, we see in Fig. \ref{fig:Rot}a that PbTiO$_3$ cannot sustain a spontaneous polarization along $z$ under large compressive strain $\eta_{zz} < -2.5$\% (i.e. the $P4mm$ curve coincides with the $P4/mmm$ curve for $\eta_{zz} < -2.5$\%) but the system does not become paraelectric: instead, it prefers to stay ferroelectric and to develop a polarization in the perpendicular direction ($Pmm2$ phase). Both under tensile and compressive strains, the polarization is typically enhanced compared to the bulk value (Fig. \ref{fig:Rot}). 

Independently, we also considered different possible phases including AFD motions. According to what was discussed for the bulk, we only considered the most favorable $a^-$ AFD motions. The $AFD_{xy}$ ($Imma$) and $AFD_{xyz}$ ($C2/c$) phases are compatible with the rotation patterns $a^-a^-c^0$ and $a^-a^-c^-$ respectively. The strain evolution of the relaxed rotation angles of both phases are shown in Fig. \ref{fig:Rot}. We observe that the relaxed $AFD_{xyz}$ phase only combines rotations along the three cartesian directions in a small region of strain, around $\eta = 0$: under tensile strain, it prefers a purely $a^0a^0c^-$ rotation pattern while, under compressive strain, it prefers a purely $a^-a^-c^0$ rotation pattern (i.e. it reduces to the $AFD_{xy}$ phase). In all cases, the gain of energy produced by the AFD motions is much smaller than what can be obtained from the polar distortion.

It is worth noticing also in Fig. \ref{fig:strain_phase} that, as at the bulk level, the polarization always prefers to stay aligned with one of the pseudo-cubic axis ($z$ or $x$) and that the $FE_{xy}$ phase is never the most stable. Nevertheless, its energy is very close to that of the $FE_x$ phase. Contrary to the latter, under compressive uniaxial strain for which lattice constants perpendicular to the constrained direction are elongated, the $FE_{xy}$ phase develops an AFD instability. This instability is associated to the $a^-a^-c^0$ AFD motions of the $AFD_{xy}$ phase which appears to be the most favorable AFD configuration under compressive strain. Condensing these additional AFD motions in the $FE_{xy}$ phase brings the system into a $FE_{xy}$+$AFD_{xy}$ phase ($a^-_Pa^-_Pc^0$ in generalized Glazer's notation) of $Ima2_1$ symmetry that is lower in energy that the purely $FE_{xy}$ phase but is however never more stable than the $FE_x$ phase. This contrasts with the prediction of a $Ima2_1$ ground state for PbTiO$_3$ under tensile epitaxial biaxial strain \cite{ericthesis}. The difference of behavior can be explained by the fact that the biaxial tensile strain forces two elongated lattice constants to be equal favoring a $FE_{xy}$+$AFD_{xy}$ distortion while, under uniaxial compressive strain, the lattice constants in the two directions perpendicular to the constraint are similarly elongated but the system keeps the freedom to break the symmetry between them. 

We see in Fig. \ref{fig:strain_phase}b that the gains of energy associated to the FE distortions are amplified and those associated to AFD motions significantly reduced with the GGA-WC in comparison to the LDA. Still the system switches from a $FE_x$ ground-state to a $FE_z$ ground state at a relatively similar critical strain $\eta_{zz} = 1.5$\%.   

In conclusion, under uniaxial strain, PbTiO$_3$ adopts a purely ferroelectric ground state independently of the strain amplitude, with the polarization aligned either along the constrained direction ($FE_z$ phase) for $\eta_{zz} < \sim +1$\% or perpendicular to it, along one of the pseudo-cubic directions ($FE_x$ phase), for $\eta_{zz} < \sim +1$\%.

 \section{Uniaxial stress}

\begin{figure}[htbp!]
\centering
\includegraphics[scale=.35]{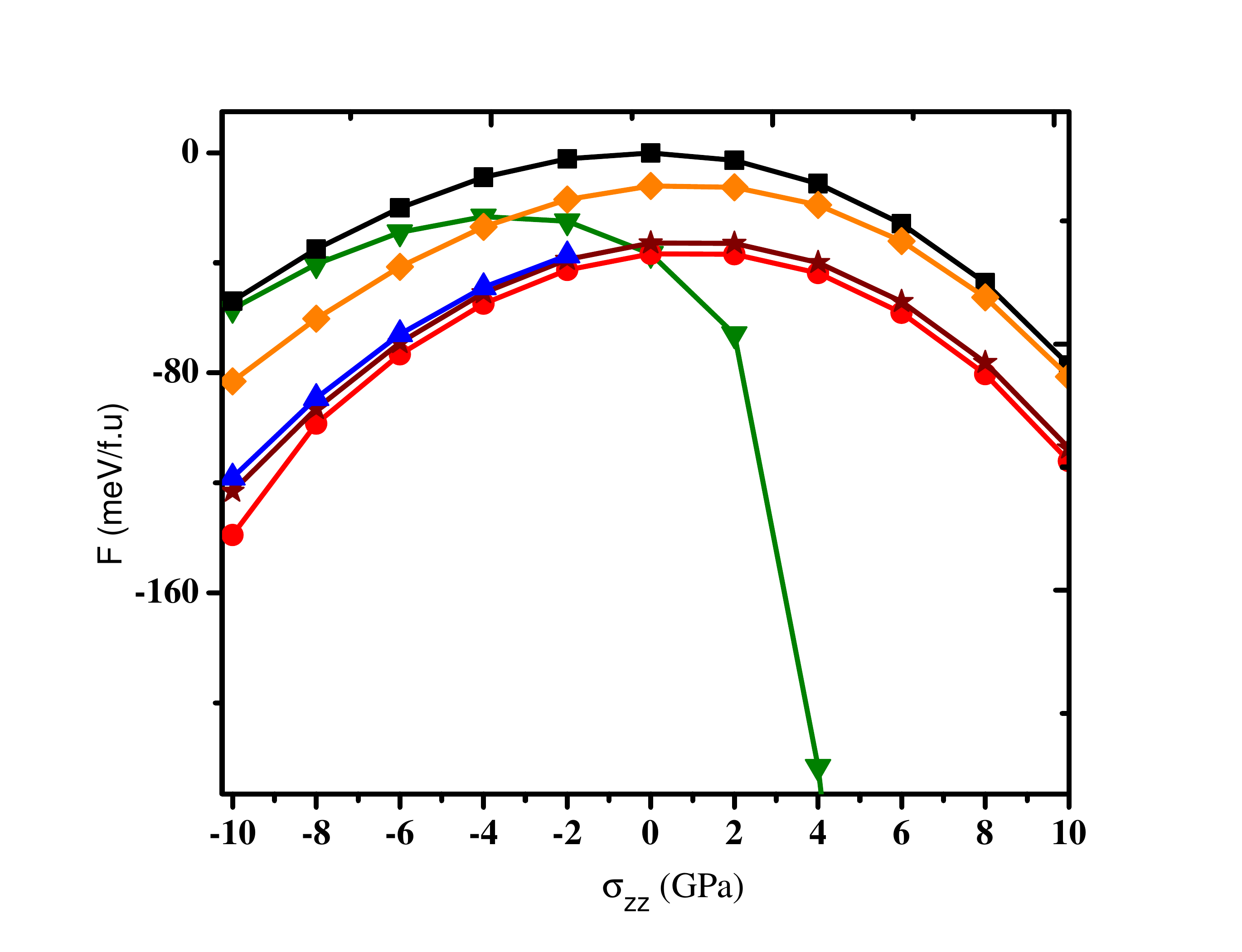}
\caption{{(Color Online) Mechanical enthalpy $F$ (meV/f.u.) of different metastable phases of PbTi$O_3$ under uniaxial stress as computed within the LDA. The considered phases are the following: PE (black squares), $P_z$ (green triangles), $P_{xy}$ (blue triangles), $P_x$ (red circles), $AFD_{xy}$ (orange diamonds) and $AFD_{xy}$+$P_{xy}$ (brown stars).}}
\label{fig:stress_phase}
\end{figure}

Since it is more easily accessible experimentally, let us now consider the behavior of PbTiO$_3$ under uniaxial stress, $\sigma_{zz}$. In this case, the stable phase is the one which minimizes the mechanical enthalpy $F = U - \sigma_{zz} \eta_{zz}$. The LDA results are summarized in Fig. \ref{fig:stress_phase}. We see that, as for fixed uniaxial strain, the ground state of PbTiO$_3$ under fixed uniaxial stress is always purely ferroelectric. At $\sigma_{zz} = 0$, the system has degenerated ground states, corresponding to having the polarization either along $z$ ($FE_z$ phase) or perpendicularly to it ($FE_x$ phase) \cite{footnote_2}. As expected, the presence of uniaxial tensile stresses always favors the $FE_z$ phase while uniaxial compressive stresses always stabilizes the $FE_x$ phase. Again, under compression, the $AFD_{xy}$+$P_{xy}$ phase appears very low in energy and below the $P_{xy}$ phase but is never more stable than the $FE_x$ phase.

\begin{figure}[htbp!]
\centering
\includegraphics[scale=.35]{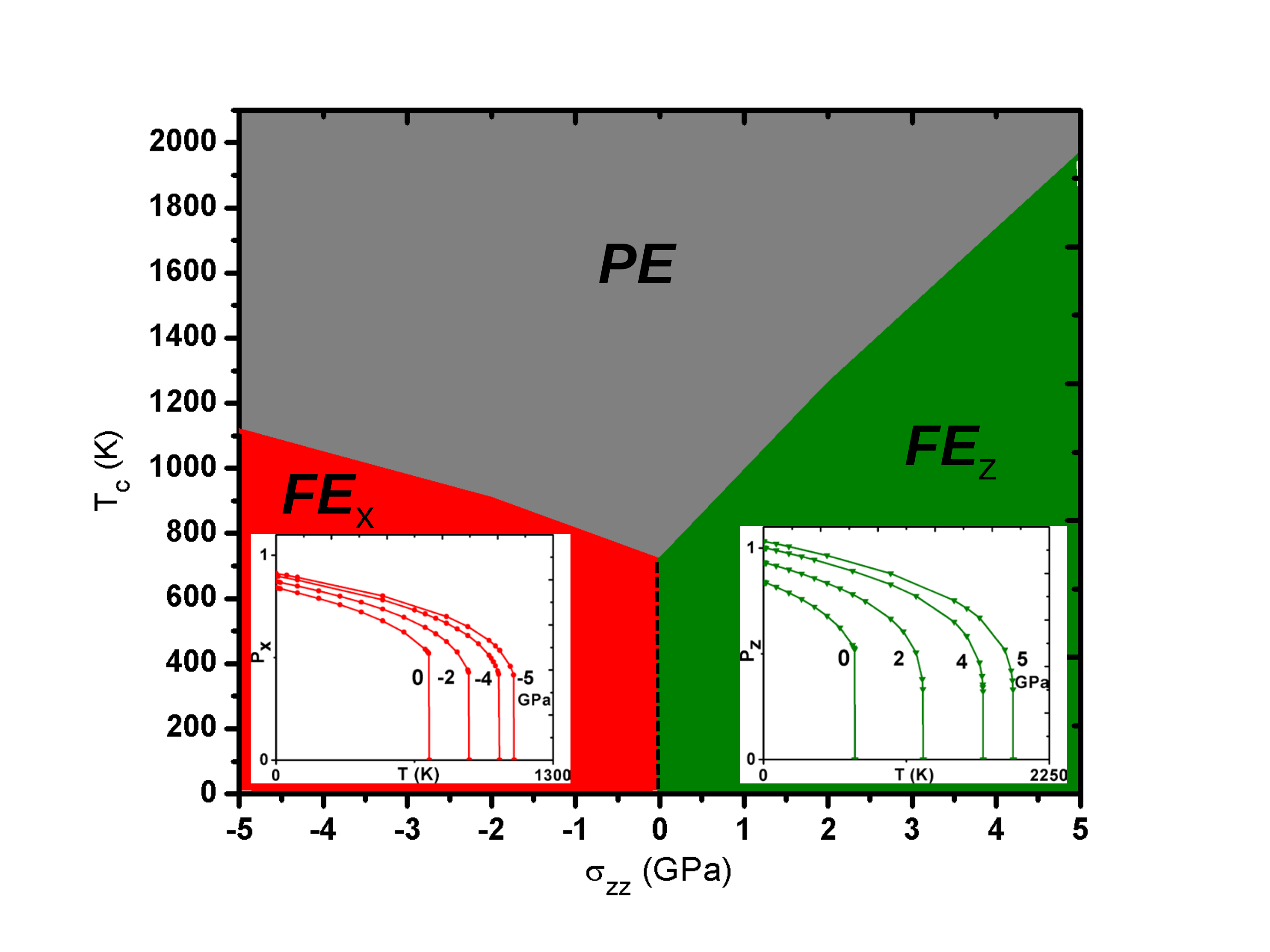}
\caption{{ (Color Online) Phase diagram of PbTiO$_3$ under uniaxial stress $\sigma_{zz}$, as predicted from LGD theory.}}
\label{fig:critical_temp}
\end{figure}

Since the AFD motions does not appear to be directly involved in the ground state, the behavior of PbTiO$_3$ under uniaxial strain can be further explored using a simple Landau-Ginzburg-Devonshire (LGD) theory, including the order parameter $P$ and neglecting the AFD degrees of freedom. The phase diagram of PbTiO$_3$ under uniaxial stress predicted from LGD model is reported in Fig. \ref{fig:critical_temp}. In our calculations, we have used the same parameters as Qiu {\it et al.} \cite{Nagaranjan}. The LGD results are in agreement with our first-principles calculations, reproducing a $FE_x$ ground state under compressive stress and a $FE_z$ ground-state under tensile stress. The uniaxial stress both increases the saturated polarization and linearly shifts the phase transition temperature to higher temperatures. 
%We see also that the phase transition that is first-order at $\sigma = 0$, tends to become second-order under uniaxial stress.  

\begin{figure}[htbp!]
\centering
\includegraphics[scale=.35]{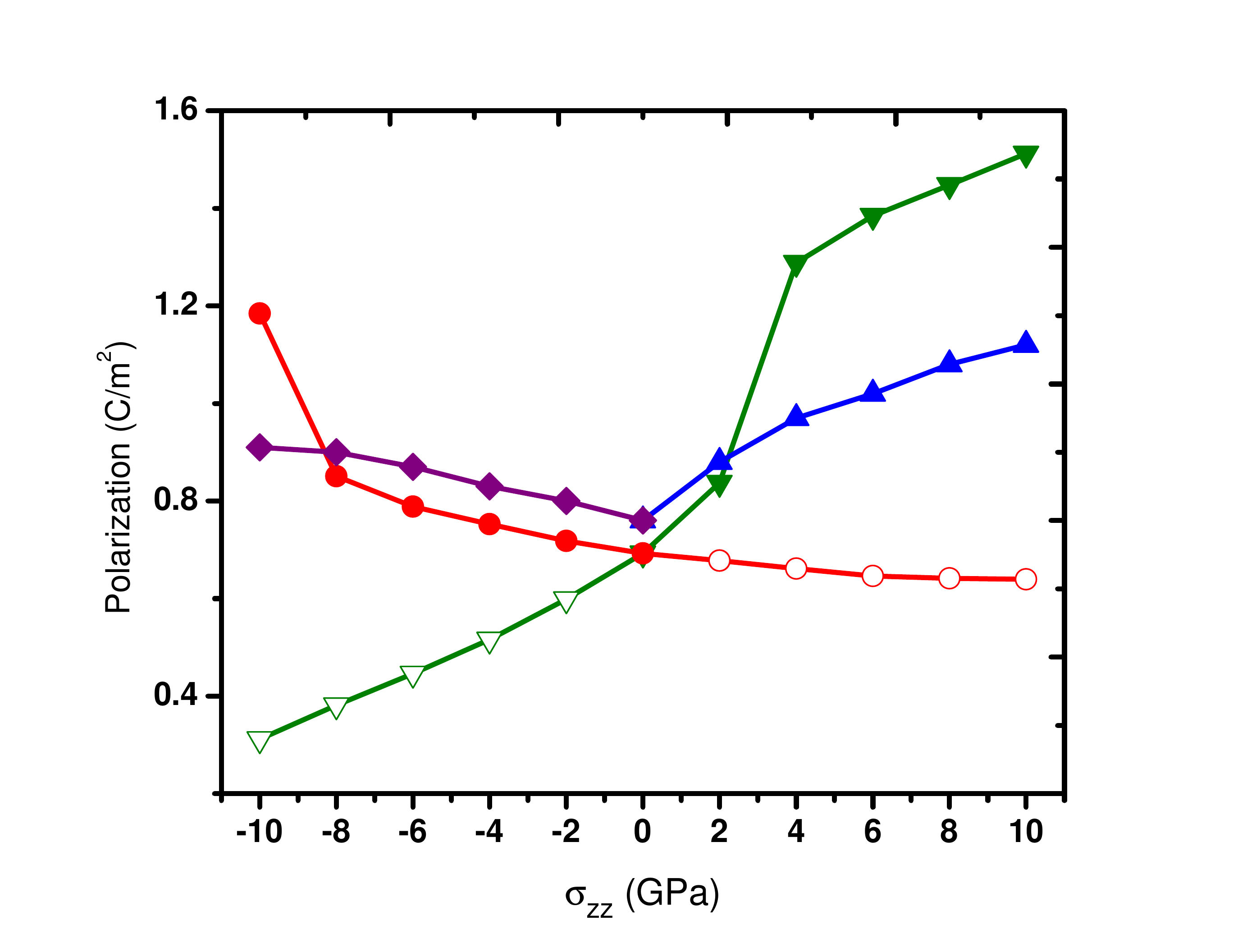}
\caption{{(Color Online) Evolution of the polarization of the $FE_x$ (red circles) and $FE_z$ (green down triangles) phases of PbTiO$_3$ with the uniaxial stress, as predicted within the LDA. Filled symbols correspond to the region where the phase is the ground-state. Purple diamonds and blue up triangles correspond to the prediction from LGD theory at 300K \cite{Note-LGD} for the $FE_x$ phase and $FE_z$ phase, respectively.}}
\label{fig:Polarization}
\end{figure}

In Fig. \ref{fig:Polarization}, we report the evolution of the spontaneous polarization $P_s$ of PbTiO$_3$, as a function of the applied uniaxial stress.  Although first-principles and LGD calculations nicely agree in a wide range of compressive stress, they only coincide in the limit of small tensile stress. The first-principles calculations reveal an abrupt jump of $P_s$ at a critical tensile stress $\sigma_{zz}^c \approx 2$ GPa that is not captured in the LGD model. This jump of the polarization of the $FE_z$ phase under tensile stress was previously highlighted by Duan {et al.} \cite{yifeng}. We see that a similar behavior also appears in the $FE_x$ phase under compression, but at much larger critical stress. 

\begin{figure}[htbp!]
\centering
\includegraphics[scale=.35]{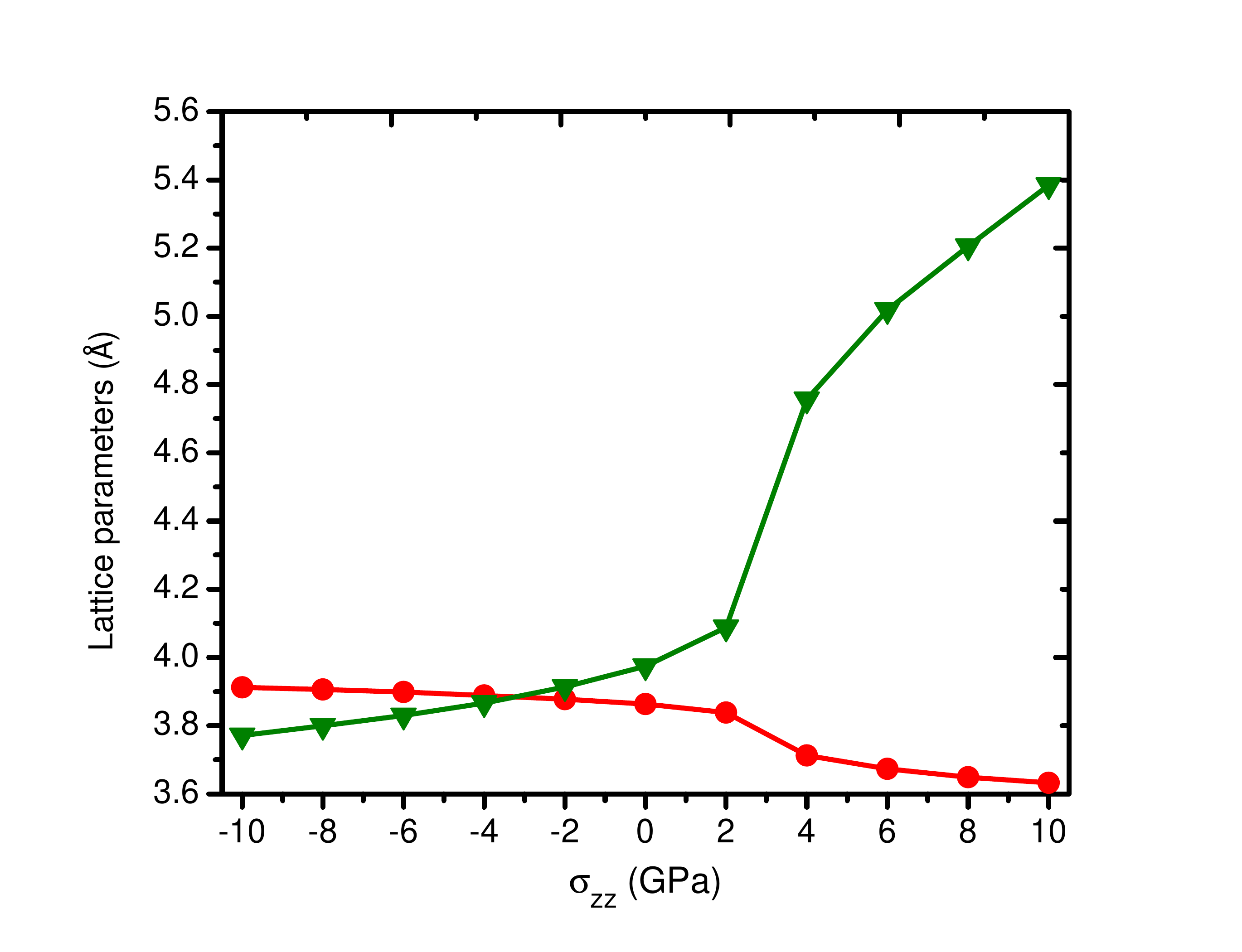}
\includegraphics[scale=.35]{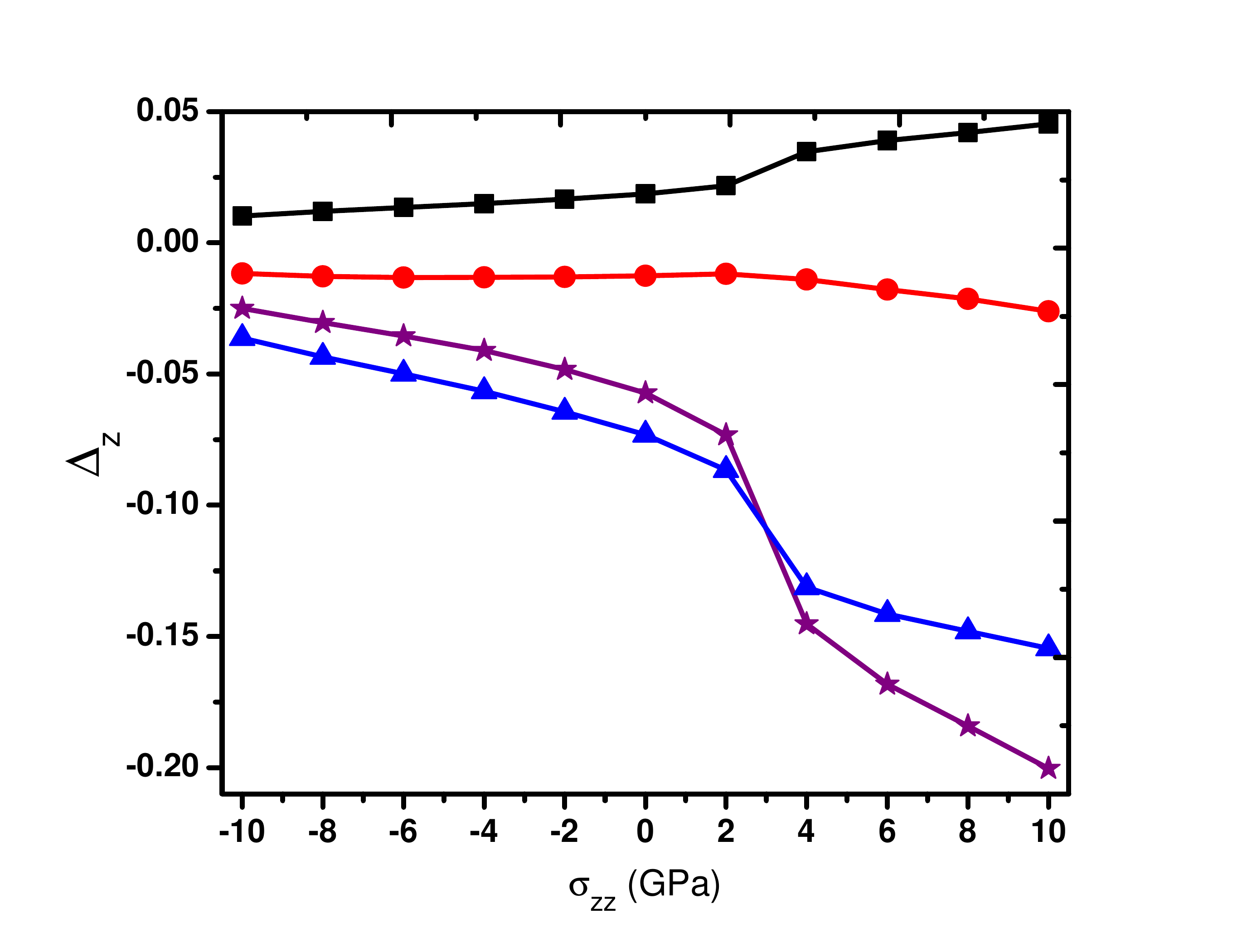}
\includegraphics[scale=.35]{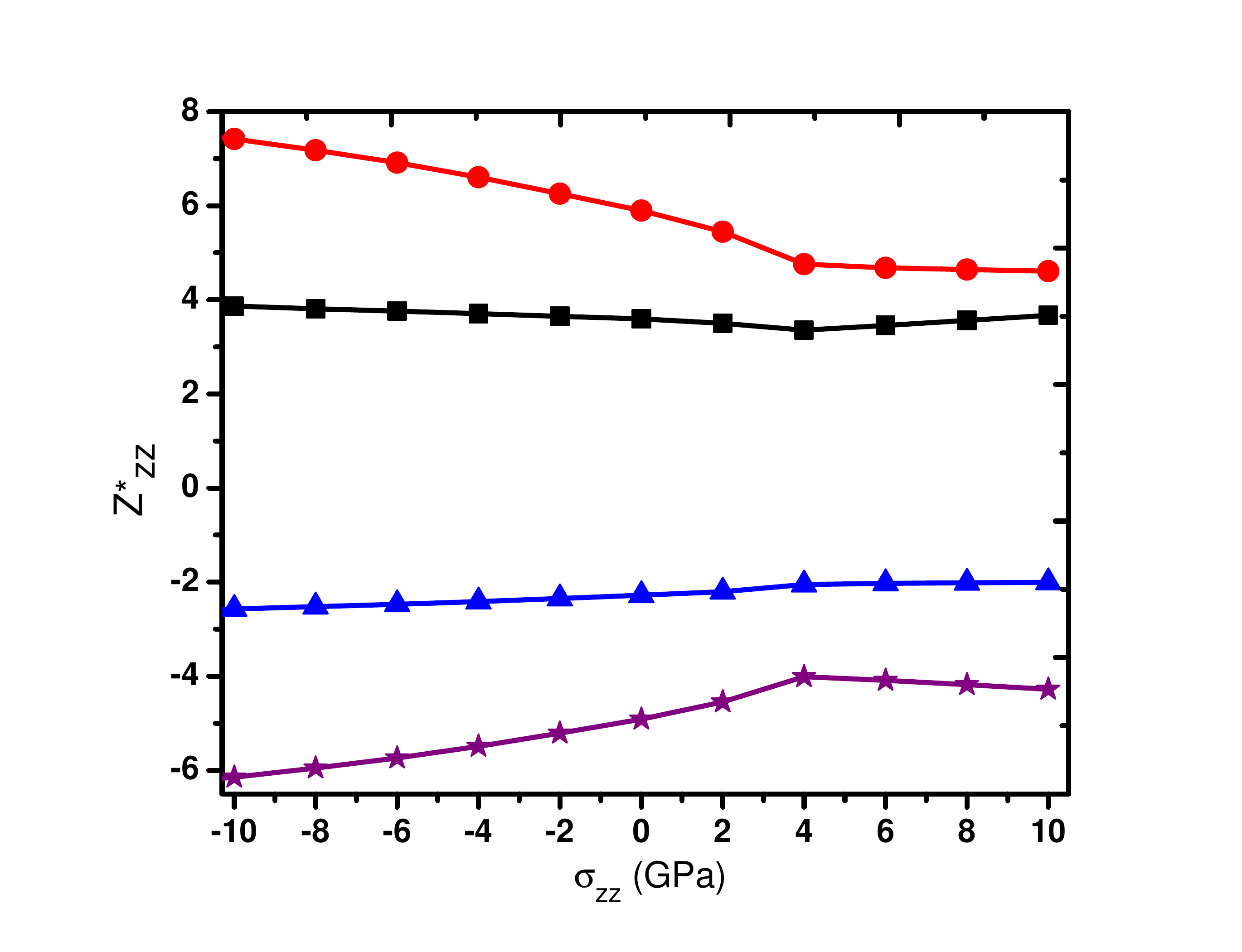}
\caption{{ (Color Online) Evolution of lattice parameters c (green triangles) and a (red circles) in  (\AA) (panel a), atomic displacements  (panel b) and Born effective charges (panel c) as a function of uniaxial stress. Panel b and c: black squares for Pb, red circles for Ti,  purple stars for  O$_1$, blue triangles for O$_{2,3}$. }}
\label{fig:SuperT}
\end{figure}

As illustrated in Fig. \ref{fig:SuperT}, the sudden increase of $P_s$ is linked to a dramatic jump in the $c$ parameter and accompanied with a strong ionic relaxation. It will only be partly compensated by a small decrease of the Born effective charges. This behavior (including the evolution of the atomic distortions) is totally comparable to what was previously reported for PbTiO$_3$ under isotropic negative pressure \cite{Tinte}. Tinte {\it et al.} explained that behavior by the proximity of a phase transition, the microscopic origin of which could be the breaking of one of the Ti--O bonds along the polar axis. 

Here, it appears however at a smaller critical tensile stress. Moreover it is predicted also under compressive stress. While negative isotropic pressure is something not practically accessible experimentally, uniaxial stresses (both tensile or compressive) were recently made accessible to lab on chip experiments \cite{Raskin,Gravier,Bhaskar}. This could at first offer the possibility to confirm our prediction experimentally. Moreover, it could also reveal of concrete practical interest : as highlighted by Duan {et al.} \cite{yifeng}, in the vicinity of the critical stress, PbTiO$_3$ will exhibit a large piezoelectric response  i.e. $d_{zzz} = \partial P_z/æ\partial\sigma_{zz}$ ($d_{xzz} = \partial P_x /æ\partial \sigma_{zz}$) and proportional to the slope of $P$ in Fig. \ref{fig:Polarization} that might be directly exploited to enlarge the piezoelectric sensitivity of nanodevices.

\section{CONCLUSIONS}

The behavior of PbTiO$_3$ under uniaxial strains and stresses has been explored from first-principles calculations and LGD theory. Under uniaxial strain, PbTiO$_3$ adopts a purely ferroelectric $FE_x$ ground state under compressive strain and switches to a purely ferroelectric $FE_z$ ground state under tensile strain larger than $\eta_{zz}^c \approx 1$\%. This contrasts with the emergence of phases combining FE and AFD distortion under biaxial strain and isotropic pressure. Under uniaxial stress, PbTiO$_3$ exhibits either a $FE_x$ ground state under compression or a $FE_z$ ground state under tension. Moreover, our calculations highlight an abrupt jump of the structural parameters under both compressive and tensile stresses at critical values $\sigma_{zz} \approx +2$ GPa and $- 8$ GPa. While LGD theory reproduces nicely the first-principles data it does not capture this strong relaxation and so remains only valid in a region between the critical stresses. The jump of the structural parameters will be linked to a strong increase of the piezoelectric response, which might be potentially exploited. We hope that our work will motivate further experimental characterization of PbTiO$_3$ under uniaxial tensile and compressive stresses.

\section*{Acknowledgments}
This work was supported by Grenoble INP funded by IDS-FunMat, an International Doctoral Programme in Functional Materials. Additional financial support has been provided by the Gabriel Lippmann Public Research Center (Luxembourg), through the National Research Fund, Luxembourg (FNR/P12/4853155/Kreisel) and the University of Li\`ege (Belgium) through the ARC project TheMoTherm. Ph.G. acknowledges a Research Professorship from the Francqui Foundation.


\begin{thebibliography}{9}
\bibitem{footnote_1}
Relaxations with a polarization developing along the 3 directions of space have also been performed but 
relaxed back to $FE_x$ or $FE_z$ phases, except in the vicinity of $\eta_{zz} = 0$ and, in that case, where 
always producing a lower gain of energy than $FE_x$ or $FE_z$ configurations.
\bibitem{Note-P}
The spontaneous polarization was also estimated multiplying the atomic displacements by the Born effective 
charges of either the reference cubic phase or of the relaxed FE phase at each stress. The Berry phase calculation
is much closer to the estimate.
\bibitem{Note-LGD}
The LGD model predicts a very similar evolution at low temperatures but with a slightly larger amplitude. 
Since LDA tends to slightly underestimate the polarization, we have made the comparison with LGD theory 
at 300K.
\end{thebibliography}
\end{document}